\documentclass[aps,showpacs,a4paper,floatfix,twocolumn,prc,amsmath,amssymb]{revtex4-1}

\usepackage{amsmath,amssymb}
\usepackage{graphicx}
\usepackage{epsfig,epstopdf}
\usepackage{ulem}
\usepackage{natbib}
\usepackage{color}
\usepackage{appendix}
\usepackage{amsfonts}
\usepackage{hyperref}

\begin{document}
\title{Brazilian physicists community diversity, equity and inclusion: a first diagnostic}
\author{Celia Anteneodo$^1$}
\affiliation{$^1$ Departamento de Física, Pont\'ificia Universidade Cat\'olica do Rio de Janeiro, \\
PUC-Rio, Rua Marqu\^es de S\~ao Vicente 225, 22451-900 Rio de Janeiro, RJ, Brazil}
\email{celia.fis@puc-rio.br}
\author{Carolina Brito$^2$}
\email{carolina.brito@ufrgs.br}
\author{Alan Alves-Brito$^2$}
\affiliation{$^2$ Instituto de F\'isica, Universidade Federal do Rio Grande do Sul, \\Av. Bento Gonçalves, 9500, 91501-970 Porto Alegre, RS,  Brazil}
\author{Simone Silva Alexandre$^3$}
\affiliation{$^3$ Departamento de F\'isica, Universidade Federal de Minas Gerais, \\Av. Ant\^onio Carlos, 6627, 31270-901 Belo Horizonte Minas Gerais, MG, Brazil}
\author{Beatriz Nattrodt D'Avila $^4$}
\author{D\'ebora Peres Menezes $^4$}
\email{debora.p.m.26@gmail.com}
\affiliation{$^4$ Departamento de F\'isica, CFM - Universidade Federal de Santa Catarina, \\ C.P. 476, CEP 88.040-900, Florian\'opolis, SC, Brazil}

\begin{abstract}
 We report the results of a survey applied to students and professionals in the area of physics in Brazil, pursuing to draw a portrait of the composition of this community in terms of  the social markers age, race, ethnicity, geographical origin, sex, gender, sexual orientation, and disabilities.   The main goal was to quantify the representativeness of different groups in the community and to detect motivations and difficulties encountered by each group throughout their studies and career. This survey was open to the members of the Brazilian Physics Society (SBF) from July to September 2018.  
 Our outcomes reveal that (i) the Brazilian physicists community is poorly diverse even in comparison with the population composition, (ii)      the main obstacle to pursue the career is socioeconomic vulnerability and (iii) harassment is high in our society, being more pronounced among women.
  We hope that these results will be useful to scientific and educational institutions to develop different strategies and policies to change this current situation.

\end{abstract}

%
\maketitle
 
\section{Introduction}

One of the greatest educators in Brazil, Paulo Freire, recognized the enriching role of diversity  for a socially committed, liberating and critical education~\cite{Freire2003}. 
Moreover, 
diversity, equity and inclusion can increase creativity and contribute to  individual and collective development~\cite{Alves-Brito2019, SAXENA201476, mckinsey2015},
besides providing equal opportunities and avoiding the loss of talented people.
 Through new concepts and perspectives, 
education and science can benefit  and, reciprocally,  exert a positive  feedback  in pro of changing the observed picture of inequality~\cite{philips2014, Malone2015, EUbook2000}.
This is specially desirable in a country like Brazil, whose economy is one of the strongest in the world and, at the same time, one of the most unequal~\cite{hdr}. 

As shown by many recent studies, Brazil is specially unequal and an extremely difficult country for women, LGBTQI+ (lesbian, gay, bisexual, transgender, queer and intersex) people, 
Afro-descendants and members of indigenous communities, as well as individuals with disabilities \cite{SINTESE,IBGE, MAPAVIOLENCIA_IPEA, UNreport}. 
As a structural problem, the Brazilian scientific environment is not an exception to the rule, 
being dominated by white~\cite{Morcele2019} and male people~\cite{Barbosa2013, Menezes2017, FERRARI2018}, implying under-representation of other groups, specially in leadership or prestigious positions,   with respect 
to the demographic composition of the full population~\cite{FERRARI2018,marcia}. 
However, no records about the Brazil's physics community exist beyond markers as age, sex and place of residence.

 In view of all that, we pursued to   draw 
a quantitative picture of the level of diversity, inclusion and equity
amongst the members of the  Brazilian Physical Society (SBF, in Portuguese). Then,  we applied a survey,  using  the social markers age, race,  ethnic, geographical origin, sex, gender, sexual orientation and disabilities.
The main goal of this paper is to present  the initial characterization  of  the Brazilian physical community that emerges from the outcomes of the survey.  
 Detecting motivations and difficulties faced by the physics community is essential to help define policies that can improve the quality of education, the formation of new human resources and, consequently, the quality of research in the field.

This paper is organized as follows: after a brief description of the historical background and of how the data were
obtained, we present the results of survey. 
They are divided into four separate subsections: the
profile of the respondents, motivation and difficulties during their
studies and along their careers, the problems of   harassment in the community, and the job market available for physicists. A summary and final remarks are then presented. The questionnaire is displayed in the Appendix.

\section{Background}
\label{background}

 Historically, Physics Education in Brazil has been the object of systematic study since the 1960s, strongly influenced by the implementation, in the United States and, soon after, in Latin America, of the Physical Science Study Committee project~\cite{PSSC}. 
Since then, Physics Education in Brazil has undergone several transformations following the different historical periods of the country (military regime of 1964-1985; democratic period after the 1988 Constitution) until today, in which there is, for education, science, technology and culture in general, a critical moment of setback and disbelief in scientific knowledge. Today, despite the advances and setbacks, the area of Physics Education in Brazil is well consolidated, presenting varied thematic lines, which are not only focused on learning, but also on historical, political, philosophical and social issues~\cite{abrapec}.

 The SBF  was founded on 14 July, 1966. 
It is a nonprofit association formed by physicists and scientists from related areas with activity or collaborations
in Brazil. Its office is located in S\~ao Paulo city. In 2003, the Gender Relations Commission was established by the SBF board of directors, aiming to verify the possible existence of gender gaps in the academic environment of physicists  and implement policies to solve this problem. Currently called Gender Working Group (GTG, in Portuguese), it presents here a critical diagnostic concerning key aspects directly related to social markers of difference faced by physicists.

\section{Data and Methods}
\label{method}

\subsection{Main questions}
 \label{questions} 
 
The main  questions that guide this study are: how diverse is the physics community? What attracts people to pursue this career? What are the main difficulties in becoming and working as a physicist? To what extent are sexual and moral harassment problems in this community? How are physicists placed in the job market? 
An answer to these questions  may give hints that  help us to build, on the basis of quantitative data, public policies for science education and outreach more sensitive to diversity, inclusion and equity.

 \subsection{Survey}
 \label{sciencequestions2}

The questionnaire presented to the SBF members 
(initially posed in Portuguese) is 
translated in the Appendix. It was built in GoogleForms to be filled anonymously, and 
was open for responses from July 3 to September 21, 2018. 
It consisted of various sets of questions, most of them in multiple choice format, to facilitate the participation of respondents as well as  data collection and analysis. A window for free comments was also available alongside most questions. 
It was  inspired mainly in a survey applied in 2018  to  students of federal institutions~\cite{IFES_survey}, 
to make a census of that universe. It was chosen as a basis (in alternative to surveys undertaken by foreign institutions)
because of the terminology being in Portuguese and  following official Brazilian census, which was particularly important concerning race and gender issues. It was adapted to our particular questions and a preliminary version of our questionnaire was tested (as a pilot survey) in a small group of about 20 people that gave us a feedback to improve the questions. It was also revised by  the members of the Working Group of Under-represented People of the SBF. Then it was launched within the whole community for volunteer anonymous participation. 
It was advertised by e-mails sent to the SBF members but, as its anonymous character was preserved, we cannot guarantee that people from outside the community have not replied.  

As said above, the terminology followed the official Brazilian census and less common terms were defined in the questionnaire.   
The Brazilian Institute of Geography and Statistics (IBGE), since the 19th century,  
is the official agency responsible to define census categories, which are based on a skin color continuum, ranging from individuals with very fair skin to those individuals with a very dark one. We have adopted IBGE official categories: {\it branca} (white), {\it preta} (black), {\it parda}, 
{\it amarela} (yellow: translated as Asian) and {\it indígena} (indigeneous). 
In Brazil, there is a common distinction between people who self-declare themselves as black, with a darker tone of skin and {\it parda}, with lighter skin tones~\cite{IBGEEduca_cor}.

After discarding a few incorrectly filled forms (less than 1 \%), the number of valid  ones was 1695, which is impressive given there   were 3875 effective  SBF members  at the time of the survey. 
From this set, multiple choice answers were counted within different groups characterized by the specific  social markers.
  
\section{Data Analyses and Results}

\subsection{Profile of the respondents}
\label{profile} 
 
In this section we address the questions about how diverse is the physics community in Brazil and how diversity changes when the career evolves. 
We first display information about the academic level attained. We then analyze diversity according to geographical origin and place of residence, race and ethnicity, sex, gender and sexual orientation, and disabilities.

\subsubsection{Academic degree}

In Table~\ref{table:profile},  we show the
composition of the population of  1695 respondents, with respect to their education level (highest attained academic degree). Let us remark that  Bachelor
  in Physics and Physics Education are both undergraduate degrees at the same level of education, and ``related
  areas'' refer to  other majors 
   in science, technology, engineering, mathematics
  (STEM).    
Notice that the majority are graduate  people, i.e., 66\% and 19\% hold a PhD and a Master degree in Physics (or related areas), respectively,
while 9\% hold a degree in Physics Education or a Bachelor degree in
Physics (or related areas), and 6\% are undergraduate students that
have only finished high school studies. 
These values are compatible with the records of  the SBF about the composition of members with annuity payment on time, also shown in  Table \ref{table:profile}.
The proportion of respondents is a bit larger amongst affiliates with doctoral degree and lower amongst undergraduate students (hence, with High School (HS) degree).

\begin{table}[h]
\centering
\begin{tabular}{|l||c| c || c |c|c|}
\hline
     {\bf Degree}        &\multicolumn{2}{|c||}{\bf Respondents}  &\multicolumn{2}{|c|}{\bf SBF Members} \\  \hline
       & number& percent &number & percent \\ \hline  \hline 
High School  &104   & 6     & 505  & 13    \\ \hline 
Bachelor     & 67 &2     &279 &5   \\
\hline
Physics Education  & 94  &7    & 59 &4   \\
\hline
Master  & 314&19   &  708  &18   \\
\hline
Doctoral & 1116&66   &  2324  &60   \\
\hline
{\bf Total} & {\bf 1695}& {\bf 100  } &  {\bf 3875}& {\bf 100  } \\
\hline
\end{tabular}
\caption{Composition of respondents according to academic degree, compared to the composition of SBF affiliates. 
(Bachelor  and Physics Education are  undergraduate degrees at the same level of education.)}
\label{table:profile}
\end{table}

\subsubsection{ Geographical aspects}

\begin{figure}[h!]
\centering
\includegraphics[width=0.85\columnwidth]{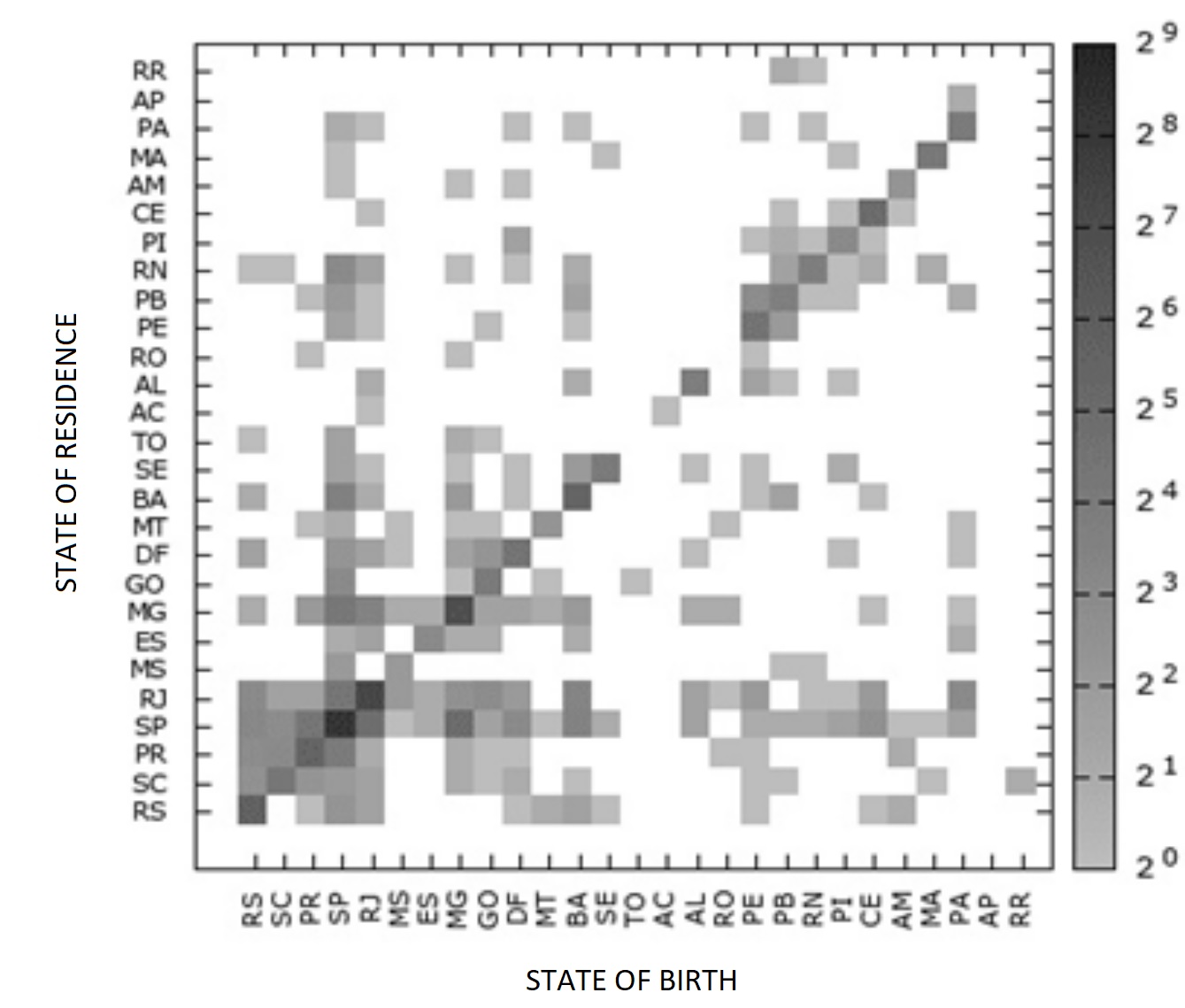}
\caption{Current state of residence versus state of birth. 
The gray scale represents the number of individuals. Brazilian states (two letter code) were ordered by latitude of their capital. 
From the respondents 1594 are Brazilian and 1659 live in Brazil.
}
\label{scatter}
\end{figure}

Concerning geographical origin (place of birth) and current place of residence, the matrix in Fig.~\ref{scatter} displays the number of people in each State of Brazil (identified by a two letter code and ordered by increasing latitude of its capital).  
The population is concentrated in the South and Southeast regions and there is also a nucleus in the Northeast including and above Pernambuco (PE). One also observes that there is little mobility of physicists in Brazil in general, probably due to availability of many universities in each State.
The matrix highlights  migration towards and from the States of São Paulo (SP) and Rio de Janeiro (RJ),  which are those with largest 
gross regional product (GRP) per capita~\cite{PIB-Brazil},  while the other two more populated States of Minas Gerais (MG) and Bahia (BA) display predominantly     migration from rather that migration to. A large mobility is also observed within the three States of the South (i.e., RS, SC and PR),  alongside SP and RJ in the Southeast region, and also within a few states in the Northeast with epicenter at PE.

\subsubsection{Race/ethnicity}
 
\begin{figure}[h!]
\centering
\includegraphics[width=0.85\columnwidth]{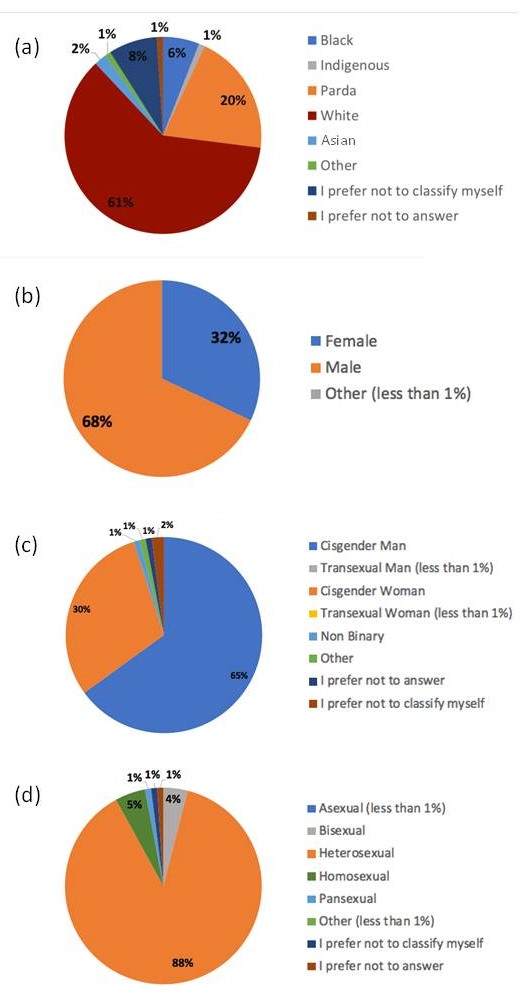}
\caption{Profile of the respondents according to  
ethnicity/race (a),  sex (b), gender (c) and  sexual orientation (d).
}
\label{fig:Profile2}
\end{figure}

The (self-declared) ethnic composition of the community is presented in Fig.~\ref{fig:Profile2}a. 
Among the respondents,  
30\% declared to be non-white (including  
  black, {\it parda},  indigenous, Asian and 2\% who declared ``other'') while 9\% preferred not to answer or classify themselves.
This information is not available in  SBF records. 
As a reference, in Brazil, black and {\it parda} people constitute 54\% of the population~\cite{IBGEEduca_cor}.  
 Apart from that, the  racial and ethnic  statistics produced by the IBGE  shows that Brazil is not a racial democracy at all; on the average, white people in Brazil have the highest salaries and face less unemployment~\cite{IBGENoticias_salarios_cor}. 

\begin{figure}[h!]
\centering
\includegraphics[width=0.9\columnwidth]{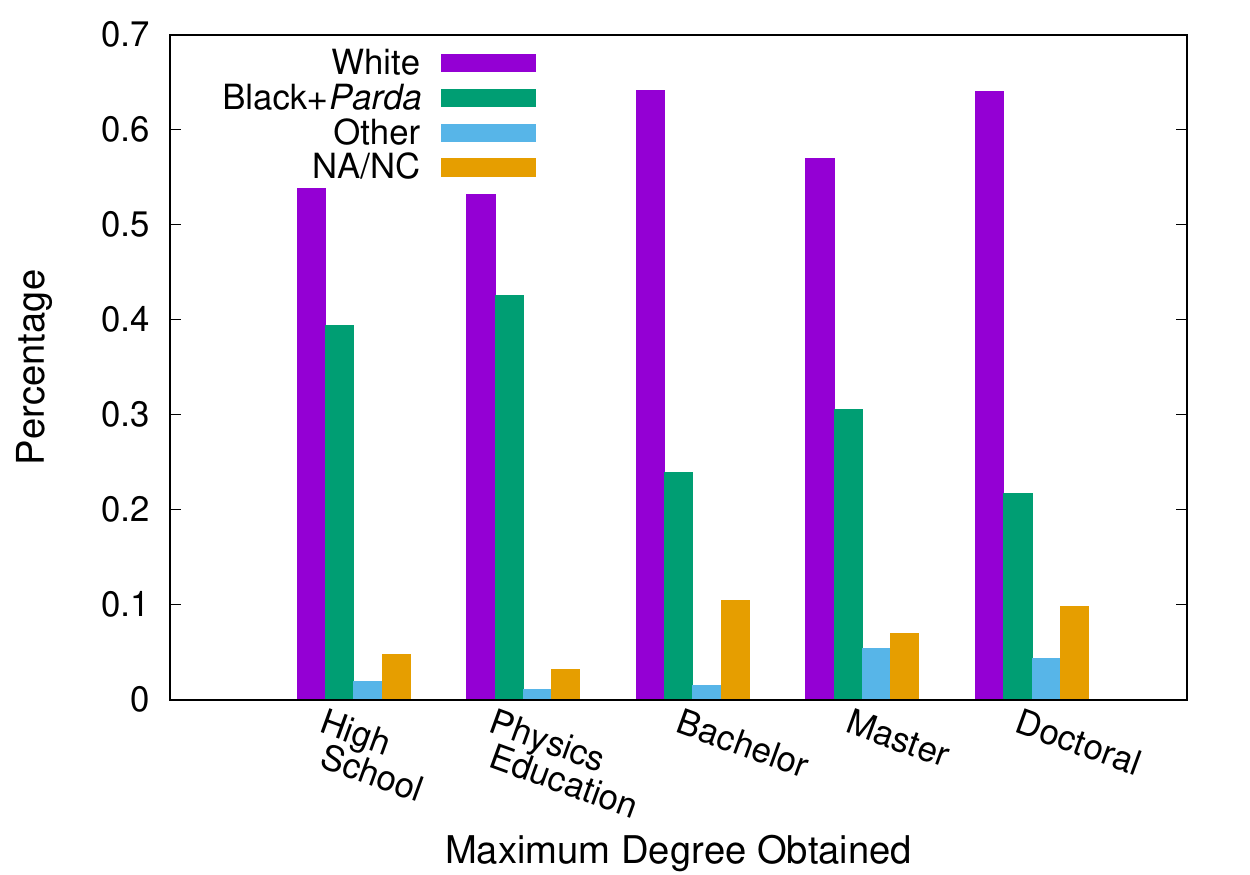}
\includegraphics[width=0.9\columnwidth]{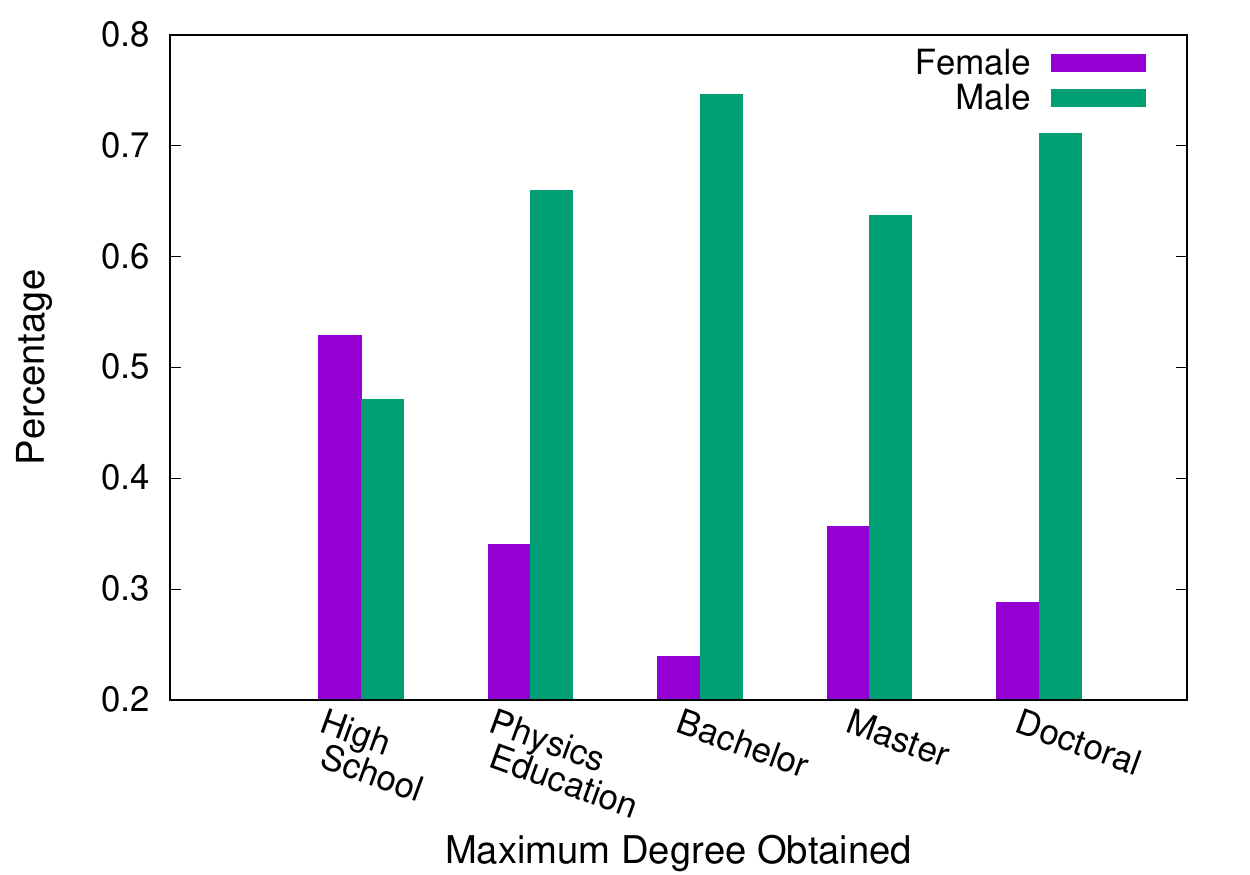}
\caption{
Percentage of respondents vs. highest degree attained, separated  by race/ethnicity group (top) and by sex (bottom).  In the legend NA/NC$=$ prefer not to answer or classify themselves.  Recall that Physics Education and Bachelor correspond to under-graduate degrees at the same level. 
}
\label{fig:tesoura1}
\end{figure}

Figure~\ref{fig:tesoura1}(top) shows the percentage of people that attained each degree decomposed by racial/ethnic groups. 
This profile puts into evidence the neat tendency of  a lower proportion of black$+${\it parda} people with the progression in the carrier.
A similar trend has been detected in the STEM field in Brazil, where black and {\it parda} people have more representation amongst students than among researchers~\cite{Morcele2019}.

\subsubsection{Sex/gender/sexual orientation}

In this section we discuss the composition with respect to sex, gender and sexual orientation.

As shown in Fig.~\ref{fig:Profile2}b, the distribution of self-declared sex among the respondents is 32\% female (slightly above 27-28\% female community members), 68\%  male and less than 1\% ``other''.  
Among the respondents, 95\%  self-declared to be cis-gender, about 3\% preferred not to answer or classify, while about 2\% self-declared trans-gender  
(see Fig.~\ref{fig:Profile2}c).
Moreover, noticeable percentages of people that answered the survey declared
themselves either as homosexual 
 or bisexual (totalizing about 10\% (see Fig.~\ref{fig:Profile2}d).
 
It is remarkable that women are the majority of respondents with a high-school degree and  among first-year undergraduate  students, however their relative presence tends to decrease amongst those that finish the undergraduate  stage  (both Bachelor and Physics Education).  
At the level of graduate studies, the same trend is observed, with decreasing proportion of women finishing Doctoral studies than Master ones. 
These profiles outline what is called ``Scissors effect''  with respect to sex, many times observed in Brazil~\cite{Barbosa2013, Menezes2017, FERRARI2018} 
and abroad~\cite{EUbook2000,reportElsevier2018,Argentine_survey}. 
Let us call the attention on  the 
similarity of this profile and that observed in the upper panel of Figure \ref{fig:tesoura1}  with respect to white and black+{\it parda}.

A noticeable feature is that there are more people from
underrepresented categories (both in the cases of sex and race) with a degree in Physics Education than with a Bachelor degree in
Physics. This may be related to the necessity to attain a degree that
allows to enter earlier in the job market, although high school
teachers in Brazil are comparatively badly paid. Another trace of
these underrepresented groups is that their proportion with Master
degree is higher than with Bachelor degree, but similar to the
proportion with degree in Physics Education, suggesting that  
 although the initial choice was to follow  the Education
carrier, they finally opted to pursue graduate studies.

In sum, the majority of the respondents were white heterosexual men, as previously observed in other scientific communities~\cite{EUbook2000,reportElsevier2018,Morcele2019}. Scientists who identify themselves as LGBT are a minority within
  the respondents. Consequently, they can feel themselves marginalised in different ways.  In addition, previous international studies have shown that the campus environment can be quite LGBTphobic, specially for transgender individuals~\cite{rankin2003,rankin2010}. It is also worth  noting that in Brazil, for instance, the life expectancy of a transgender individual (measured in 2013) was less than half that of the Brazilian population~\cite{trans-idade,trans-report}.

\subsubsection{Disabilities} 
When asked about disabilities, 1590 declared none. That is, only 6.2\% of the respondents declared to   bear a  disability. 
Amongst them, the most common (not excluding) answers were: low or abnormal vision (61 answers), physical and motor limitations (27), global developmental disorder (7), and amongst ``others'' the most frequent case was bipolar disorders (7). Among those who declared to have some disability, we found 27\% women, a bit below the population proportion.  
The percentage with disabilities agrees with that found within the Brazilian population, according to IBGE~\cite{ibge-def}, when asking about some type of deficiency 
(auditory, visual, physical or intellectual). Like in the Physics community, more than half of the cases corresponds to visual deficiencies.
 Although not shown in the tables,  41\% within the group bearing some type of deficiency declared that they reached the highest degree in their career as fast as other colleagues. A more detailed investigation of the difficulties associated with specific disabilities should be made, in order to diagnose whether the Brazilian institutions are prepared to receive students, researchers and staff with disabilities. The survey shows that they tend to progress only slightly slower than the other respondents.

\subsection{Motivations}
\label{motivations} 

The survey also aimed to identify the motivations for choosing a career in physics.  This section is devoted to present the related data.

\begin{figure}[h!]
\centering
\includegraphics[width=0.85\columnwidth]{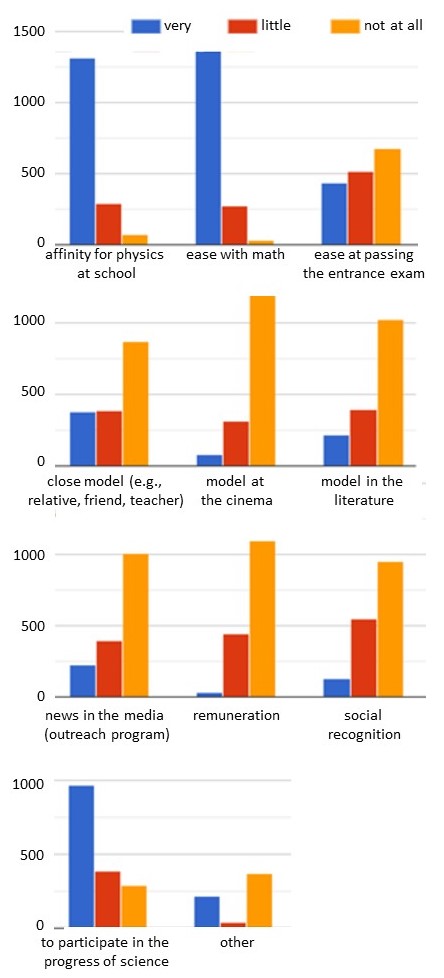}
\caption{Motivations for choosing physics. Histograms showing the number of (non-exclusive) responses associated to each
possible motivation given in a list, according to the level of identification (very, little, not at all). 
Horizontal lines are drawn as references. 
}
\label{fig:motivacao}
\end{figure}

The respondents were asked to indicate the level of  identification (very much, little or not at all) with several given possible motivations for choosing physics.
The alternatives presented in the list were not excluding, so that multiple choices could be selected.

Figure~\ref{fig:motivacao}  shows the histograms of the number of responses,  for each motivation and level of identification. The three main motivating factors are: 
(i) ease in mathematics at school, 
(ii) affinity for physics at school, and (iii) desire to take part in the progress of science.  
A high score obtained in mathematics at school was previously identified as an indicator of pursuing careers in STEM field with high probability~\cite{TaiScience2006}.
Socioeconomic aspects, such as future salaries and social recognition, rarely appear as career  main motivations.
It was also asked whether models of scientists in the literature, cinema or family environment provided a motivation to pursue a career in physics. In most cases the answer was negative, as can be seen in Fig.~\ref{fig:motivacao}, perhaps due to the absence of positive models. 

 The respondents could also indicate other possibilities not listed in the survey. 
Only 231 respondents indicated at least one 
spontaneous (not induced) motivation. 
The more popular ones are:
curiosity (33), teaching (30), knowledge (30), understanding (29), passion/pleasure(16), challenge (14), job opportunity (12), routine freedom (7).

\subsection{Detecting difficulties}
\label{difficulties} 

One of the main  purposes of the survey was to identify 
obstacles along the studies and carrier. 
Concerning this issue,  respondents were asked to identify  (in a given list) the factors that negatively influenced the studies or the carrier,  as well as the factors identified as cause of discrimination felt by the respondents in the environments of study or work. The respondents were also requested  to indicate the presumed causes of discrimination suffered by colleagues in the study and work environments,  whenever  observed or taken knowledge of.

\begin{figure}[h!]
\centering
\includegraphics[width=1.\columnwidth]{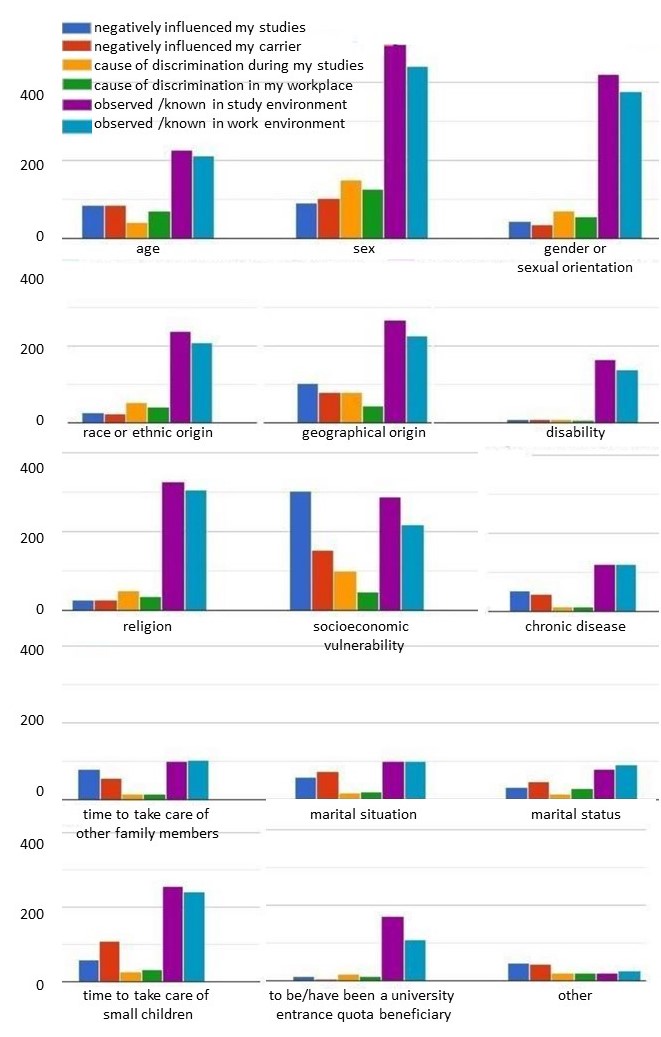}
\caption{Causes of  difficulties. 
Number of (non-exclusive) responses for  each factor 
associated to negative influence or discrimination of the respondent or colleagues given in the caption. 
Horizontal lines are drawn as references.
}
\label{fig:dificuldades}
\end{figure}

Figure~\ref{fig:dificuldades} shows the histograms of number of (non-excluding) answers. It puts into evidence that the main difficulty is related to the  socioeconomic origin. 
Even in the subgroup of those who marked ``Other'', there are various specifications related to social or economic issues. As an example: necessity of work, family needs, flaws in the education, living far away from the university.

It is noteworthy that a large percentage of respondents have already observed or are aware of discrimination of colleagues in the study  (lilac bar) and  work  (light blue bar) environments, with respect to diverse issues:  sex, gender or sexual orientation, race or ethnicity, socioeconomic status, geographical origin, child care and religion. 
This stands in contrast with the own perception of the interviewees of feeling discriminated,
which may be related to the fact that those who suffer more are underrepresented in the community, hence in the population of respondents.

In fact, for instance, amongst women, 46.7 \% indicated a negative influence or discrimination due to sex, 
while that percentage falls to only 1.2 \% for men.
In Table~\ref{table:difficulties}, we present details for the more numerous categories of race/ethnicity, 
where we show the percentage of people in each group that identified some of the listed causes as having a negative 
impact. It is evident that {\it parda} people feel a more negative impact than white people with respect to the listed possible causes, a portrait which is even worst for black people. 

\begin{table}[h]
\centering
\begin{tabular}{|l||c| c |c| c|}
\hline          &\multicolumn{4}{|c|}{\bf Causes} \\
\hline 
       & Race/color & Socio-econ. &  Religious &  Geographic \\
\hline\hline
White        & 0.4 \%  & 18.6 \%   & 4.6 \% & 0.5 \%\\
\hline
{\it Parda}  & 7.7 \%  & 32.9 \%   & 7.9 \% & 28.8 \%\\
\hline
Black        & 45.5 \%  & 49.4 \%   & 14.3 \% & 31.2 \%\\
\hline
\end{tabular}
\caption{Percentage of people (within each  group) that felt negative influence or discrimination due to the 
causes in the first row.}
\label{table:difficulties}
\end{table}

\subsection{Harassment}
\label{harassment} 

As one of the main difficulties encountered, the issue of harassment deserves a  careful and separate treatment. This section is devoted to discuss 
to which extent harassment affect the community and which are the most attained groups.

\begin{figure}[h!]
\centering
\includegraphics[width=0.95\columnwidth]{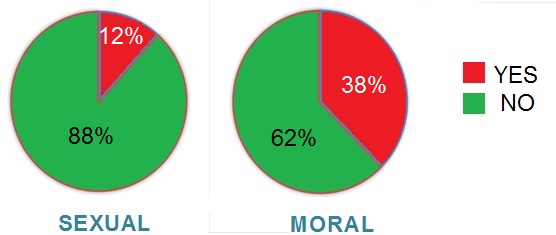}
\caption{Percentage  of respondents reporting sexual or moral harassment.}
\label{fig:asedio}
\end{figure}

The general percentages of respondents  reporting to be victim of sexual or moral harassment are summarized in Fig.~\ref{fig:asedio}. 
Moral harassment means workplace or school bullying, in the sense described in Ref.~\cite{bullying-definition} 
The questions in the survey were restricted to vertical descendent harassment, 
in order to restrict the questions specific situations more easily identifiable. 
Almost 12\% of respondents reported having suffered sexual harassment, while astonishing 38\% felt to have been victim of moral harassment.

We also analyze the percentage of Yes and No responses within each category  of race/ethnicity and sex/gender, in order to identify the more vulnerable ones.
Table~\ref{table:sexual} exhibits the characterization of the universe
of those who reported having suffered (or not) sexual harassment (12\%
in the total population). The percentage of Yes answers is much higher
within the female group (32\%) than in the male group (2\%). 
Comparing the different subgroups of the female universe, the answers are rather
homogeneous, except for the  Asian women.
A recent review  summarizes the numbers of sexual harassment found in previous studies, showing that the percentage ranges from 30\% up to 70\%  among women, in different fields and at different levels of the career~\cite{sexHass_Tenbrunsel2018}, indicating  that this is a serious problem in many other communities.

\begin{table}[h]
\centering
\begin{tabular}{|l||c| c |c||c| c |c|}
\hline 
         &\multicolumn{6}{|c|}{\bf Sex} \\\hline 
         &\multicolumn{3}{|c||}{Female}
         &\multicolumn{3}{|c|}{Male} \\ \hline
{\bf  Race/Ethnicity  }      
& Yes& No &\%Yes& Yes& No &\%Yes \\ \hline  \hline 
Black & 14 & 23 & 38 & 2 & 59 & 3 \\ \hline
Indigenous& 1 & 3 & 25 & 0 & 6 & 0 \\ \hline
\it{Parda} & 28 & 60 & 32 & 6 & 241 & 2 \\ \hline
White & 114 & 233 & 33 & 17 & 678 & 2 \\ \hline
Asian & 16 & 6 & 0 & 18 & 0 \\ \hline
Other & 3 & 5 & 38 & 1 & 15 & 6 \\ \hline
Prefer not to classify  & 8 & 22 & 27 & 2 & 95 & 2 \\ \hline
Prefer not to answer & 0 & 3 & 0 & 0 & 15 & 0 \\ \hline
{\bf Total} & {\bf 169} & {\bf 365} & {\bf 32} & {\bf 28} & {\bf 1127} & {\bf 2}\\ \hline
\end{tabular}
\caption{Number and percent of people who reported having suffered {\bf sexual} harassment, within the subgroups of race and sex. Composition of respondents according to academic degree, compared to the composition of SBF affiliates.}
\label{table:sexual}
\end{table}

Table~\ref{table:moral} shows the corresponding values of the case of moral harassment, where 38\% of respondents answered Yes. In this case, the numbers are alarming also in the male group (31\%), but among females, the numbers are even higher (52\%). In all subgroups of the female universe, the percentages of Yes answers are higher than in the whole population. In the male universe they are smaller except in the black and ``other'' subgroups.

\begin{table}[h]
\centering
\begin{tabular}{|l||c| c |c||c| c |c|}
\hline 
\hline 
         &\multicolumn{6}{|c|}{\bf Sex} \\\hline 
 &\multicolumn{3}{|c||}{Female}          &\multicolumn{3}{|c|}{Male} \\ \hline
{\bf Ethnicity/Color}
          & Yes& No &\%Yes& Yes& No &\%Yes \\ \hline  \hline 
Black & 19 & 18 & 51 & 25 & 36 & 41 \\ \hline
Indigenous& 3 & 1 & 75 & 1 & 5 & 17 \\ \hline
\it{Parda} & 42 & 46 & 48 & 81 & 166&33 \\ \hline
White & 180 & 167 & 52 & 202 & 493 & 29 \\ \hline
Asian & 7 & 10 & 41 & 2& 16 & 11 \\ \hline
Other & 4 & 4 & 50 & 10 & 6 & 63 \\ \hline
Prefer not to classify  & 19& 11 & 63 & 33 & 64 & 34 \\ \hline
Prefer not to answer & 2 & 1 & 67 & 3 & 12 & 20 \\ \hline
{\bf Total} & {\bf 276} & {\bf 258} & {\bf 52} & {\bf 357} & {\bf 798} & {\bf 31}\\ \hline
\end{tabular}
\caption{Number and percentage of people who reported having suffered
  {\bf moral} harassment (bullying), 
within the subgroups of race and sex.}
\label{table:moral}
\end{table}

\begin{table}[h]
\centering
\begin{tabular}{|l||c|c|c|| c| c |c|}
\hline 
    &\multicolumn{6}{|c||}{\bf Harassment} \\ \hline
    {\bf Sex, Gender, Orientation  }    
 &\multicolumn{3}{|c||}{Sexual}
         &\multicolumn{3}{|c|}{Moral} \\ \hline
     
 & Yes& No &\%Yes& Yes& No &\%Yes \\ \hline \hline  
Female,  Cis, All  & 161 & 351 & 31 & 265 & 247 & 52 \\ \hline
Female, Non binary, All &3 & 3 & 50 & 6 & 0 & 100 \\ \hline 
Male, Cis, All  & 25 & 1073 & 2 & 334 & 764&30 \\ \hline
Male, Other, All & 1 & 6 & 14 & 4 & 3&57  \\ \hline  
Male, Not to classify, All  & 1 & 24 & 4& 9 & 16 & 36 \\ \hline
Male, Not to answer, All & 1 & 18 & 5 & 8 & 11 & 42 \\  \hline  
Female, Cis, Heterosexual  & 134 & 310& 30 & 219 & 225 & 50 \\ \hline
 Female, Cis, Bisexual& 15 & 19 & 44 & 19 & 15 & 56 \\ \hline
Female, Cis, Homosexual & 4 & 14 & 22 & 11 & 7 & 61\\ \hline
Female, Cis, Not to classify & 5 & 5 & 50 & 6 & 4 & 60\\ \hline 
Male, Cis, Heterosexual& 21 & 971 & 2 & 294 & 698 & 30\\ \hline
Male, Cis, Bisexual& 0 & 22 & 0 & 9 & 13 & 38\\ \hline
Male, Cis, Homosexual& 4 & 62 & 6 & 28 & 38 & 42\\ \hline
\end{tabular}
\caption{Sexual and moral harassment by category of sex, gender identity  and sexual orientation. Only categories with at least one response are shown.  "All" refers to all categories of sexual orientation. 
}
\label{table:sexualandmoral}
\end{table}

In Table~\ref{table:sexualandmoral}, the numbers represent the total of respondents about sexual and moral harassment in each subgroup of gender and sexual orientation. Only cases with more than five answers are shown. This table 
also puts into evidence the higher percentages of sexual and moral harassment in the non-cisgender population. Although the absolute numbers of non-cis populations are not very large, some aspects are striking. This is the case, for example, of the non-binary female population, where all the six women report having suffered bullying and half of them sexual harassment. In the case of the male population, four of the seven respondents in the ``other'' category reported having suffered bullying.  This proportion of 57\% is almost twice that observed in the male population of cisgender men.
The percentages of moral and sexual harassment are systematically higher in the non-heterosexual (both female and male) populations, except for case of  cisgender-homosexuals.
In general, the data point to a higher incidence of moral and sexual harassment in populations that differ from the majority response profile regarding gender identity (cisgender majority) and sexual orientation (heterosexual majority).
This trend is similar to that reported in a research about undergraduate students from six universities in Canada regarding their experience of sexual violence~\cite{Storey_JAH2018}.

\subsection{Job market}
 \label{market} 

Next analysis refers to the insertion of the physicists community in the labor market. 

The question about professional occupation allowed multiple
choices. This enables a combination of various responses, including
high school (HS) teacher, university researcher or lecturer (faculty, in the USA), temporary job (substitute [up to two years in Brazil], hourly, etc.),
permanent job, position in a company or industry, autonomous, scholarship holder, or ``without work or scholarship''. 
Let us remark that public universities in Brazil are free of charges and tuition fees.  Scholarship holders are students who earn a certain amount of support money for basic subsistence costs during their studies.  

\begin{figure}[h!]
\centering
\includegraphics[width=0.95\columnwidth]{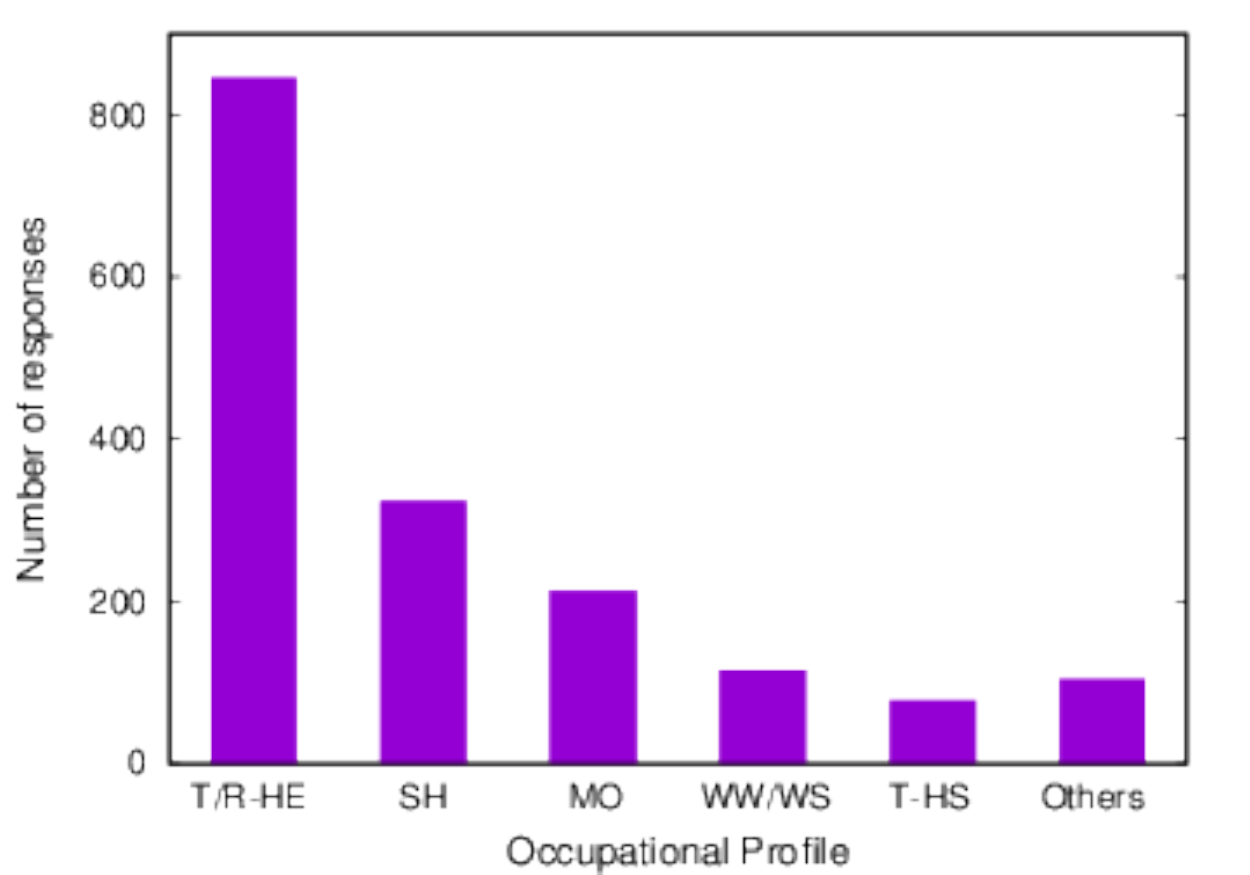}
\caption{Distribution of occupation of the respondents. The initials
  in the abscissa axis correspond to the following categories:  {\bf
    T/R-HE} = lecturers and/or researchers in  higher education
  institutions, {\bf SH} = scholarship holders, {\bf MO} = multiple
  occupations, {\bf WW/WS} = without work nor scholarship, {\bf T-HS}
  = high school teachers. }
\label{fig:ocupacao}
\end{figure}

\begin{table}[h]
\centering
\begin{tabular}{|l|l||c|c|}
\hline 
\multicolumn{2}{|l||}{\bf Occupation} 
 &\multicolumn{2}{|c|}{\bf Number}
        \\ \hline \hline
T/R-HE & & & 845\\ \hline
SH & & & 323 \\   \hline
T-HS & & & 77\\ \hline 
Other & & &103\\   
& Company or industry (CI)    & 24 &  \\
& Autonomous (A)                  &  7 &  \\
& HS substitute                   & 16 &  \\
& Temporary (T)  & 26 &  \\
& Permanent               & 30 &  \\
\hline  
MO & & &                213 \\
& HE+HS                         &138 & \\
& HE+HS+SH+A           &11 & \\
& SH+(T or A)  &17 & \\
& T+(CI or A)  &4 & \\
\hline 
WW/WS & & &115\\ \hline \hline
{\bf Total} & & & {\bf 1676} \\ \hline 
\end{tabular}
\caption{ 
Occupational profile of respondents: further details. 
{\bf T/R-HE} = lecturers and/or researchers in  higher education
institutions, {\bf SH} = scholarship holders, {\bf MO} = multiple
occupations, {\bf WW/WS} = without work nor scholarship, {\bf T-HS} =
high school teachers.  
}
\label{tab:ocupacao}
\end{table}

Table~\ref{tab:ocupacao} summarizes the occupational profile of
respondents of the questionnaire. Almost half of the answers were from
people who are lecturers/faculty and researchers in higher education. 
 Of those who reported having  multiple occupations, the most common combination refers to people who are high school (HS) teachers and university lecturers simultaneously. 
The numbers are given in Table.~\ref{tab:ocupacao},
which also presents further details, with subdivisions for the cases
``multiple occupation'' and ``other''. Noticeably, in this universe of
responses we find a high percentage of people who accumulate
functions: 12\% of physicists work in more than one professional
activity. Among them, there are 138 individuals who are HS teachers
and university faculty, of which 50 specify that they are substitute teachers in HS. This high percentage of people who accumulate positions deserve attention, and  is plausibly related to underpaid positions.
We also investigated the profile of unemployed people in the physics community and verified that, among the 115 respondents in the subgroup of "without work or scholarship", 43 have HS as highest level of education and 40 have a Doctoral degree as highest level. Normalising these numbers by the total respondents in each one of these categories, given in Table~\ref{table:profile}, we conclude that only $3.6\%$ of the total respondents with a doctorate are unemployed, while around $41\%$ of the undergrad  students do not have scholarship. 
 
We now take a closer look at the  subgroup of people working in higher education. This subgroup is numerically representative (almost half of the total) and is considered a privileged group in the universe of SBF members. The reason is that the career of higher education lecturers offers higher salaries than the HS teacher career so that people do not have to accumulate activities as declared by more than 12\% of the community and already discussed above. 
In this universe, 95\% has a doctorate, 29\% are female, which is slightly lower than the percentage of female respondents (which is 32\%).

In the clipping of race/ethnicity, there is a clear privilege of self-declared white respondents over self-declared {\it parda} and black respondents: 67\% of the respondents are white and only $16\%$ {\it parda} and  $4\%$  black.

When the question of children is evaluated, there is an important difference between the universe of female and male respondents. In addition to the direct data presented in the table that indicates that  male with children, who are in higher education is 15\% higher than the total of respondents, there is another important aspect: male and female with at least one child represents, respectively, 65\% and  44\% of the total respondents. This difference suggests that female lecturers or researchers in higher education institution needed or chose to give up having children more than male physicists did.

\section{Summary and Final Remarks}
\label{conclusions}

In this paper we have presented the results of a survey applied in the community of physicists who are members of the Brazilian Physical Society (SBF), with the main goals of addressing the following questions: How diverse is the physics community? What does attract people to follow this career? Which are the main difficulties to become a physicist? To what extent is  sexual and moral harassment a problem in this community? How are physicists placed in the labor market?  
In this section we highlight some of our findings.

The first striking point is that the Brazilian Physical Society is not diverse in any way, as shown in Sec.~\ref{profile}: it is made up of men (68\%), white (61\%), heterosexual (88\%) and southeastern (59\%) people,  in a country where black$+${\it parda} people, as well as women, are majority~\cite{IBGEEduca_cor,IBGEEduca_sex}. Moreover the diversity decreases with the progression in the career: percentages of women and black$+${\it parda} people is higher among under-graduated students than at the PhD level, as shown in Fig.~\ref{fig:tesoura1}.

We have asked about the motivations to study physics and identified that there are two main drives to 
pursue this career (Sec.~\ref{motivations}): the ability to handle math and physics at school strongly influences this decision for $68\%$ of the respondents, and around $62\%$ of the respondents point to the  hope of contributing to the progress of science. 
Approximately $80\%$ of people attribute little or 
no influence of ``social or financial recognition'' for choosing a career in physics. These results  point to a lack of recognition of scientific careers in the Brazilian society.   

Concerning the difficulties in career advancement (see  Sec.~\ref{difficulties}),
socioeconomic  problems were identified as the main 
obstacles at the beginning of the career. 
Other factors received a low number of answers but this seems to be associated to the fact that they are important only for minorities, as shown in Table~\ref{table:difficulties}. 
The  social class issue is one of the most important aspects to be considered in order to understand the lack of diversity in  Physics and the segregation of our society.  In the basic education, more than 80\% of the Brazilians attend public schools~\cite{inep_matriculas}, which has numerous deficiencies. Because private schools are expensive and the places in the public universities are limited to the ones who are able to succeed in a competitive test, people from middle and lower social classes, when they succeed entering the  university -- usually through quotas that  exist for  a decade only -- they face  enormous difficulties. 
On the one hand, follow-up policies are lacking in the universities and, on the other hand, the discourse of meritocracy, which assumes equal opportunities, is very strong in the physics community.
 
A worrying result (presented in Sec.~\ref{harassment}) concerns the very high percentage of moral harassment in the physics community: 38\% of the respondents reported to have suffered moral harassment 
 and it is more prevalent among women, being reported by 52\% of the female respondents and by 31\% of male respondents. 
Concerning sexual harassment, the total percentage is 12\%.
In this case the gender difference is more striking: 32\% of the female respondents reported being victim of a sexual harassment while only 2\% of men reported this problem. 
 
In general, we observe a higher incidence of moral and sexual harassment
in subgroups away from the majority response profile regarding
gender identity (cisgender majority) and sexual orientation
(heterosexual majority). These expressive percentages are in line with
recent international reports~\cite{NAP24994,French_survey,GenderGapinScience}.

As discussed in Sec.~\ref{motivations}, 
about 6.2\% of the respondents declared to be carriers of
specific disabilities. 
Further studies and more specific surveys should be carried out to investigate more deeply other aspects possibly neglected in the present work. 

Let us mention that recent studies have identified an extremely high concentration of mental problems in academia: in~\cite{Evans2018},  the authors reported that graduate students are more than six times as likely to experience depression and anxiety as compared to the general population. Moreover, they found that both transgender/gender-nonconforming and female graduate students are significantly more likely to experience anxiety and depression than their male graduate student counterparts.
The incidence of these mental problems depending on antecedents of sexual and moral harassment in academia remains to be investigated, but certainly these results are an alert for the institutions. 

Looking for literature about  problems faced by the community, it is worth to mention a survey~\cite{problems} that puts into evidence a source of unsatisfaction nowadays, which is the pressure to publish and its bad consequences.

Last, our survey allowed us to investigate some characteristics of the labor market for physicists in Brazil, as shown in Sec.~\ref{market}. We have identified that  most of the respondents who have a PhD in physics are employed (see Table~\ref{tab:ocupacao}): only  3.5\% of doctors who responded are out of work. On the other hand, 12\% of physicists work in more than one place and also there are 37\% of undergraduate students without scholarships.  By studying in detail the population who is unemployed or without scholarship, a bias in terms of color or gender was not found. However, we did identify that 60\% of men working in higher education have children, while only 40\% of women do. This expressive difference suggests that more women had to give up motherhood rather than man had to give up fatherhood, which is similar to the data reported in Ref.~\cite{ecklund11}.

 The present study brings results that imply not only thinking about historical and philosophical aspects of science education, but also to help us reflect on the potential that diversity develops in the production of valid knowledge. 
Although we recognise that some effort has been made in Brazil to increase the number of underrepresented people in academic environments over the last 15 years, extra commitments are still required in dealing with the social exclusion and apartheid system in sciences, specifically in physics, where the numbers reveal high levels of inequality. We hope that the results presented and discussed in this paper can serve for our community to be aware of the current scenario and for policymakers to take decisions towards the improvement of diversity, equity and inclusion,  
making the environments for Physics learning and research  more motivating and healthy, increasing the feeling of belonging to underrepresented groups. 

The complete report at the SBF website can be found in~\cite{relatorio} and is available in Portuguese only. Its translation to English is 
presented in the Appendix.

{\bf Acknowledgments:} The authors thank all anonymous respondents for
answering the survey and the SBF board and staff support.

\section*{Appendix - Questionnaire} 

``The  Working Groups of the Brazilian Physical Society (SBF in Portuguese) about Gender and Underrepresented groups have made efforts  towards a more inclusive
and egalitarian physics community in Brazil. In order to know better the diversity of the
physics community in what concerns age, race, ethnic origin, sexual
orientation and special need issues, we invite you to answer this questionnaire.
Our aim is to collect information that can contribute to the promotion
of actions and policies that help decrease the access, career and
promotion barriers, specially of those historically underrepresented
groups in our community. All answers will be treated with anonymity.

We thank your participation.''

\vspace{0.5cm}

{ \bf \it DEMOGRAPHIC CHARACTERIZATION}

\begin{itemize}
\item Year of birth (4 digits):
\item Brazilian State (2 letters) or country of birth (3 letters):
\item Brazilian State (2 letters) or foreign country (3 letters) in
  which you currently reside:
\item Do you have children? How many?
\begin{itemize}
\item I do not have
\item Yes, 1
\item Yes, 2
\item Yes, 3
\item Yes, 4 or more
\end{itemize}
\item If you have children, when were they born (in relation to the
  doctoral period)?
\begin{itemize}
\item Before
\item During
\item After
\end{itemize}
\end{itemize}

{\bf Academic training}
\begin{itemize}
\item What is the maximum academic level you have completed?
\begin{itemize}
\item High school
\item Graduation
\item Bachelor's degree
\item Master's degree
\item Doctorate degree
\end{itemize}

\item Year of completion of the highest level (4 digits):
\end{itemize}

{\bf Professional occupation}
\begin{itemize}
\item Indicate all options that apply:
\begin{itemize}
\item High school teacher
\item Higher education researcher or  lecturer/faculty 
\item Temporary job (temporary teacher, hourly, etc.)
\item Effective job
\item Work in company or industry
\item I work autonomously
\item Scholarship holder (types of scholarship: initiation to
  research, Master's, Doctorate and post-doc)
\item Without work or scholarship
\end{itemize}
\end{itemize}

{\bf Color or race}
\begin{itemize}
\item  What is your color or race?
\begin{itemize}
\item  Asian 
\item White
\item Indigenous
\item{\it Parda}
\item Black
\item Other
\item I prefer not to classify myself
\item I prefer not to answer
\end{itemize}
\item If you have answered ``other'', please specify.
\end{itemize}

{\bf Sex and gender}
\begin{itemize}
\item What is your sex?
\begin{itemize}
\item Female
\item Male
\item Other
\end{itemize}
\item If you have answered ``other'', please specify.
\item What is your gender identity?
\begin{itemize}
\item Cisgender woman (1)
\item Cisgender man (1)
\item Transsexual /Transgender woman (2)
\item Transsexual / transgender man (2)
\item Non-binary (3)
\item Other
\item I prefer not to classify myself
\item I prefer not answer
\end{itemize}
\item If you have answered ``other'', please specify.

(1) If you identify yourself with the sex assigned  at birth.
(2) If you have another gender identity other than that assigned at birth.
(3) If you do not define your gender identity within the binary system man - woman.

\end{itemize}

{\bf  Sexual orientation}
\begin{itemize}
\item What is your sexual orientation?
\begin{itemize}
\item Heterosexual
\item Homosexual
\item Bisexual
\item Pansexual
\item Asexual
\item Other
\item I prefer not to classify myself
\item I prefer not to answer
\end{itemize}
\item If you have answered ``other'', please specify.
\end{itemize}

{\bf  Disabilities}
\begin{itemize}
\item Do you have any disability? Which one?
\begin{itemize}
\item No
\item Low or abnormal vision
\item Blindness
\item Deafness
\item Physical
\item Intellectual
\item Global Developmental Disorder (2)
\item Other
\end{itemize}
(2) Autism, Rett Syndrome, Heller Syndrome, Asperger's Syndrome or Global
development without further specification.
\end{itemize}

{\bf \it MOTIVATIONS}

{\bf Indicate how much you identify yourself with the following motivations to choose the area of physics:}
\begin{itemize}
\item Affinity for physics at school
\begin{itemize}
\item Very
\item Little
\item Not at all
\end{itemize}
\item Easiness for math at school 
\begin{itemize}
\item Very
\item Little
\item Not at all
\end{itemize}
\item Ease at passing the entrance exam
\begin{itemize}
\item Very
\item Little
\item Not at all
\end{itemize}
\item Close model (e.g., relative, friend, teacher)
\begin{itemize}
\item Very
\item Little
\item Not at all
\end{itemize}
\item Model at the cinema
\begin{itemize}
\item Very
\item Little
\item Not at all
\end{itemize}
\item Model in the literature
\begin{itemize}
\item Very
\item Little
\item Not at all
\end{itemize}
\item News in the media (outreach program)
\begin{itemize}
\item Very
\item Little
\item Not at all
\end{itemize}
\item Remuneration
\begin{itemize}
\item Very
\item Little
\item Not at all
\end{itemize}
\item Social Recognition
\begin{itemize}
\item Very
\item Little
\item Not at all
\end{itemize}
\item To participate in the progress of science
\begin{itemize}
\item Very
\item Little
\item Not at all
\end{itemize}
\item Other
\begin{itemize}
\item Very
\item Little
\item Not at all
\end{itemize}
\item If you have answered “other”, please specify.
\end{itemize}

{\bf \it DETECTING DIFFICULTIES}

\begin{itemize}
\item In comparison with your colleagues, how fast did you advance in
  your studies to reach the maximum level you have accomplished?
\begin{itemize}
\item Faster
\item Same pace
\item Slower
\item I do not know
\end{itemize}

\item In case your advancement was slower than the others, what
do you think it was its main cause ? 
\begin{itemize}
\item Age
\item Sex
\item Gender or sex orientation
\item Race or ethnic origin 
\item Socioeconomic vulnerability 
\item Disability 
\item Chronic disease
\item Time to take care of small children
\item Time to take care of other family members
\item Marital situation
\item Does no apply 
\end{itemize}

\item In comparison with your colleagues with the same degree, how fast did you advance in
  your professional career ?
\begin{itemize}
\item Faster
\item Same pace
\item Slower
\item I do not know
\item Does no apply 
\end{itemize}

\item From the following reasons:
\begin{itemize}
\item  influenced my studies negatively; 
\item influenced my career negatively; 
\item  it was a discrimination motive during my studies; 
\item  it was a discrimination motive at my employment; 
\item  I have already observed or have been aware of the discrimination of
colleagues on the studies environment; 
\item  I have already observed or have been aware of the discrimination of
colleagues on the  employment environment, 
\end{itemize}
indicate all the options that apply to you:
\begin{itemize}
\item Age
\item Sex
\item Gender or sex orientation
\item Race or ethnic origin  
\item Geographical origin 
\item Disability  
\item Religion
\item Socioeconomic vulnerability  
\item Chronic disease  
\item Time to take care of other family members  
\item Marital situation
\item Marital status
\item Time to take care of small children  
\item To be/have been a university entrance quota beneficiary  
\item other
\end{itemize}
\item If you have answered “other”, please specify.

\item In cases of discrimination, was there any support from the
  colleagues or the Institution to solve the problem?
\begin{itemize}
\item Never
\item Yes, in some cases
\item Always
\item I do not know
\item Does not apply
\end{itemize}

\item Do you think that the role played by the Institutions in the
  case of discrimination should be more active than it currently is?
\begin{itemize}
\item Yes
\item No
\item Perhaps
\end{itemize}

\item Have you ever suffered sexual harassment from someone
  hierarchically superior than you (boss, supervisor, teacher) ?
\begin{itemize}
\item Yes
\item No
\end{itemize}
\item In cases of sexual  harassment, please explain what happened.

\item Have you ever suffered moral harassment from someone
  hierarchically superior than you (boss, supervisor, teacher) ?
\begin{itemize}
\item Yes
\item No
\end{itemize}
\item In cases of moral harassment, please explain what happened.

\item Please, give suggestions or information that you believe to be
  relevant and were not contemplated in this survey. 
\end{itemize}

$\newline$

\end{document}